\documentstyle{article}
\newcommand{\bibit}{\em}

\newcommand{\be}{\begin{equation}}
\newcommand{\ee}{\end{equation}}
\newcommand{\bear}{\begin{eqnarray}}
\newcommand{\ear}{\end{eqnarray}}

\newcommand{\D}{{\cal D}}

\begin{document}
\begin{center}
{{\large \bf
 WEYL-INVARIANT QUANTIZATION OF \\  \vspace{2mm}
                              LIOUVILLE FIELD THEORY}}\\
\vspace{3mm}
 M.\, REUTER \\
\vspace{1mm}
{\it Deutsches Elektronen-Synchrotron DESY   \\
                  Notkestrasse 85, D-22603 Hamburg, Germany }
\end{center}
\vspace{1mm}
\abstract{
Liouville field theory is quantized
by means of a Wilsonian effective action and its associated
exact renormalization
group equation.
For $c~<~1$, an approximate solution of this equation is obtained
by truncating the space
of all action functionals. The Ward identities resulting from the Weyl
invariance of the theory are used in order
to select a specific universality class for the renormalization
group trajectory. It is found to connect two conformal
field theories with central charges $25-c$ and $26-c$, respectively.
            }

\section{Introduction}

Liouville field theory \cite{can,dk,ger}
is not only an interesting topic in its own right it is also at the
heart of 2-dimensional quantum gravity and of noncritical string
theory \cite{polya,kpz}.
In the following we shall discuss the quantization of
Liouville theory within the
framework of the exact renormalization group approach
\cite{wil}. More precisely, we are going to 
use a formulation in terms of the effective average action
\cite{wet} - \cite{cs} which
is a modification
of the standard effective action $\Gamma$ with a built-in
infrared cutoff $k$. 
It is the effective action appropriate for fields which have
been averaged over  spacetime volumes of size $k^{-1}$. Stated differently,
$\Gamma_k$ obtains from the classical action $S$ by integrating
out only the field modes with momenta larger than $k$, but
excluding the modes with momenta smaller than $k$. In the
limit $k\to0$, $\Gamma_k$ approaches the standard effective action,
$\Gamma_{k\to0}=\Gamma$. Here $\Gamma$ is the generating functional
for the 1PI Green's functions and includes all quantum fluctuations.
The dependence of $\Gamma_k$ on the scale $k$ is described by an exact
renormalization group (RG) equation \cite{wet,gau}.
The quantization of a theory in terms of the effective average action
proceeds in two steps: (i) specify the
``short-distance'' or ``microscopic'' action $\Gamma_\Lambda$
as the initial value of the RG evolution at some
large scale $k=\Lambda$, and (ii) solve the RG equation for the
trajectory $\Gamma_k, k\in[0,\Lambda]$.  
(For a general discussion of the effective average action 
we refer to a recent review \cite{cop} and to a more detailed 
account \cite{rwli} of the material presented here
\footnote{To appear in the Proceedings of {\it Renormalization Group '96},
          Dubna, Russia, August 1996}.)

To start with,
let us briefly recall some basic facts about Liouville
theory which we shall need later on. We start from an arbitrary
conformal field theory of central charge $c$ coupled
to gravity.
Integrating out the matter field
fluctuations
leads to the induced gravity action \cite{polya}
\be\label{1.2}
\Gamma_{\rm ind}[g]=\frac{c}{96\pi}I[g]+\lambda\int d^2x\sqrt g, \ \
I[g] \equiv  \int d^2x\sqrt gR\Delta_g^{-1}R\ee
In a second step one
has to integrate over the metric $g_{\mu\nu}$. We shall do
this in the conformal gauge by picking a reference metric $\hat g
_{\mu\nu}$ and writing
$
g_{\mu\nu}(x)=e^{2\phi(x)}\hat g_{\mu\nu}(x)$.
Inserting this into (\ref{1.2}) and
performing the integration over the ghost fields one is
led to
\be\label{1.6}
S[\phi;\hat g]=-\frac{\kappa^2}{32\pi}I[\hat g]+S_L[\phi;\hat g]\ee
with the Liouville action
\be\label{1.7}
S_L[\phi;\hat g]=\frac{\kappa^2}{8\pi}\int d^2x\sqrt{\hat g}\left\{
\hat D_\mu\phi\hat D^\mu\phi+\hat R\phi+\frac{m^2}
{2}e^{2\phi}\right\}\ee
and
$
3\kappa^2\equiv 26-c $.
Here $\hat D_\mu$ and $\hat R$ are constructed from $\hat g_{\mu\nu}$.
The action $S[\phi;\hat g]$ equals $\Gamma_{\rm ind}[e^{2\phi}
\hat g]$ with $c$ replaced  by $c-26$. This substitution takes care
of the Faddeev-Popov determinant related to the conformal gauge
fixing. We shall assume that
$m^2\equiv 16\pi\lambda/\kappa^2 > 0$ and $\kappa^2>0$.

In the following we consider (\ref{1.7}) the classical
action of a field  $\phi$ in a fixed background geometry $\hat g
_{\mu\nu}$, and we try to quantize $\phi$ using the exact
evolution equation. Because the decomposition of the metric is
invariant under the Weyl transformation
$
\phi'(x)=\phi(x)-\sigma(x),\  \hat g_{\mu\nu}'(x)=e^{2\sigma(x)}
\hat g_{\mu\nu}(x)$,  
the quantization should respect this ``background split symmetry''.
This means that the functional integral over $\phi$ has to be performed
with the Weyl invariant measure \cite{dk}
based upon the distance function
$
ds^2_{\rm Weyl}=\int d^2x\sqrt{\hat g}\ e^{2\phi(x)}\ \delta\phi(x)^2$.
This measure is different from the familiar translation-invariant
measure which is associated to
$
ds^2_{\rm trans}=\int d^2x\sqrt{\hat g}\ \delta\phi(x)^2$.

\section{The Weyl -- Ward Identities}
The derivation of the exact RG equation \cite{wet,gau} starts
from the scale-dependent generating functional
\bear\label{2.1}
e^{W_k[J;g]}&=&
          \int {\cal D}_g\chi\exp\Big[-S_L [\chi;g]+\int d^2x\sqrt g
J(x)\chi(x)\nonumber\\
&&-\frac{1}{2}\int d^2x\sqrt g\chi(x)R_k(-D^2)\chi(x)  \Big]  \ear
where $\chi(x^\mu)$ is the ``microscopic'' Liouville field.
Here ${\D}_g\chi$ stands for either the Weyl or the translation
invariant measure.
(We omit all hats from now on.)
The last term in the square bracket of
(\ref{2.1}) is a diffeomorphism-invariant infrared cutoff.
The function $R_k(p^2)$ vanishes if the eigenvalues $p^2$ of
$-D^2=-D_\mu D^\mu$ are much larger than the cutoff $k$, and
it becomes a constant proportional to $k^2$ for $p^2\ll k^2$. The
precise shape of the function $R_k(p^2)$ is not important,
except that it has to interpolate monotonically between
$R_k(\infty)=0$ and $R_k(0)=Z_kk^2$~for some constant
$Z_k$.
Expanding
$\chi(x)$ in terms of eigenfunctions of $D^2$, we see that
in $W_k$ of (\ref{2.1}) the high-frequency
modes with covariant momenta
$p^2\gg k^2$ are integrated out without any suppression
whereas the low-frequency modes with $p^2\ll k^2$ are
suppressed by a smooth, mass-type cutoff term $\sim k^2\chi^2$.
It will be convenient to write the cutoff as
$
R_k(-D^2)=Z_kk^2C(-D^2/k^2)$
where $C$ is a dimensionless function with $C(0)
=1$ and $C(\infty)=0$.
The effective average action $\Gamma_k[\phi;g]$ is defined
as the Legendre transform of $W_k[J;g]$ with the infrared
cutoff subtracted \cite{wet,gau}:
\be\label{2.2}
\Gamma_k[\phi;g]=\int d^2x\sqrt g\phi(x)J(x)-W_k[J;g]-
\frac{1}{2}\int d^2x\sqrt g\phi(x)R_k(-D^2)\phi(x)\ee
Here $J=J[\phi]$ has to be obtained by inverting the relation
\be\label{2.3}
\phi(x)\equiv<\chi(x)>=[g(x)]^{-1/2}
\frac{\delta W_k[J;g]}{\delta J(x)}\ee
Obviously
$\Gamma_{k=0}=\Gamma$
is the usual effective
action because for $k=0$~%
the function $R_k(p^2)$ vanishes for all
$p^2$.
The limit
$\lim_{k\to\infty}\Gamma_k[\phi;g]$ is more subtle.
If $R_k$ is consistent with all symmetries of the theory
one can show that
$\Gamma_{k\to\infty}
=S$ holds true modulo a change of the bare parameters contained in
$S$, which is of no importance usually.
In Liouville theory the situation
is more involved because the cutoff is not Weyl
invariant. In order
to arrive at a quantum theory which is in the right universality
class, namely that of conformal field theories with central
charge $26-c$, some parameters in $\Gamma_{k\to\infty}$ have to
be fine-tuned to a particular value.
The starting point of the
evolution, $\Gamma_\Lambda$, is the only place
where the difference between the Weyl and the translation invariant
measure enters. In fact,
for both the Weyl and the translation invariant measure
eqs. (\ref{2.1}) and (\ref{2.2}) give rise to the
same functional RG equation (``exact evolution equation''):
\be\label{2.5}
\frac{\partial}{\partial t}\Gamma_k[\phi;g]=\frac{1}{2}
{\rm Tr}\left[\left(\Gamma_k^{(2)}[\phi;g]+R_k(-D^2)\right)^{-1}
\frac{\partial}{\partial t}R_k(-D^2)\right]\ee
Here $t\equiv \ln k$ is the renormalization group
``time'' and $\Gamma_k^{(2)}$
is essentially
the Hessian of $\Gamma_k$~with respect to $\phi$.

Our tool for the determination of $\Gamma_\Lambda$ are the Ward
identities related to the Weyl 
transformations. To 
derive them, we apply the 
transformation                                                        
$J'=e^{-2\sigma}J,\ g'_{\mu\nu}=e^{+2\sigma} g_{\mu\nu}, \ %
\chi'=\chi-\sigma        $
to the functional integral (\ref{2.1}) and use that
\be\label{2.10}
S_L[\chi';g']=S_L[\chi;g]-\frac{\kappa^2}{8\pi}
\int d^2x\sqrt g\{D_\mu\sigma D^\mu\sigma+R\sigma\}
\ee
The measure responds according to
\be\label{2.11}
{\cal D}_{g'}\chi'={\cal D}_g\chi\exp\left[\frac{\tau}
{24\pi}\int d^2x\sqrt g\{D_\mu\sigma D^\mu\sigma
+R\sigma\}\right]\ee
where $\tau=0\ (\tau=1)$ for the Weyl (translation) invariant measure.
The resulting Ward identity for $W_k$~can be rewritten as a statement
about $\Gamma_k$. It reads
\be\label{2.22}
{\cal L} \ \Gamma_k[\phi;g,\psi_i]=-\frac{26-c+\tau}{24\pi}R(x)+
{\cal A}_k(x)\ee
where
\be\label{2.23}
{\cal L}\ \equiv \ 2\frac{g_{\mu\nu}(x)}{\sqrt g}\frac{\delta}
{\delta g_{\mu\nu}(x)}-\frac{1}{\sqrt g}\frac{\delta}
{\delta\phi(x)}-2\sum_i\Delta^0_i\ \frac{\psi_i(x)}{\sqrt g}
\frac{\delta}{\delta\psi_i(x)}\ee
and
\be\label{2.24}
{\cal A}_k(x)\ \equiv\  <x|R_k(\Gamma_k^{(2)}+R_k)^{-1}|x>+{\rm Tr}
\left[\hat R_k(x)(\Gamma_k^{(2)}+R_k)^{-1}\right]\ee
with
\be\label{2.14}
\hat R_k(y) \equiv \frac{g_{\mu\nu}(y)}{\sqrt{g(y)}}
\frac{\delta}{\delta g_{\mu\nu}(y)}R_k(-D^2[g_{\mu\nu}(\cdot )])\ee
We have not yet explained the meaning of the external fields
$\psi_i(x)$.
In the derivation of the Ward identity we actually replaced
$S_L$~in (\ref{2.1})
by $S_L+S_\psi$ with
\be\label{2.20}
S_\psi[\chi;g,\psi_i]=\sum_i\int d^2x\sqrt g \ \psi_i(x)\ \exp[2(1-
\Delta^0_i)\chi]\ee
The action $S_\psi$ is Weyl invariant provided
$
\psi_i'=e^{-2\Delta^0_i\sigma}\psi_i$.
The
$\psi_i$'s could be some spin-0 primary fields of the underlying
conformal theory of matter and ghost fields. Their scaling
dimensions in absence of gravity
are $(\Delta_i^0,\Delta^0_i)$. We shall investigate
how these dimensions change when the system is coupled to
quantized gravity (``gravitational dressing'' \cite{kpz,dk} ).

The Ward identity
is an equation for the trace of the energy momentum
tensor derived from $\Gamma_k$.
When an arbitrary conformal field
theory with action $S$ and central charge $c[S]$ is
coupled to gravity the trace of its energy momentum
tensor
$
T^{\mu\nu}[S]  \equiv  2 g^{-1/2}
      \delta S / \delta g_{\mu\nu}    $
is (at least on shell) given by
$
T^\mu_\mu[S]=- c[S] R /24\pi + const $.
The classical Liouville action (\ref{1.7}) satisfies this
condition with the central charge
$
c[S_L]=3\kappa^2=26-c$. Coming back to the Ward identity,
$R_k$ vanishes for $k\to0$ so that ${\cal A}_k=0$~in this limit.
Thus (\ref{2.22}) with $\phi$~on shell and $\psi_i\equiv 0$~
tells us that the evolution ends at a theory which
is conformally invariant and has central charge
\be\label{2.19}
c[\Gamma_{k\to0}]=26-c+\tau\ee
For the Weyl-invariant measure $(\tau=0)$, this is the
correct value. 

\section{Truncation and Initial Value}
While the evolution equation is exact, it is usually not
possible to find exact solutions. Approximate solutions can be found
by the method of truncations which is of a nonperturbative
nature and does not require any small expansion parameter.
The idea is to
project the renormalization group flow on a subspace of the space
of all action functionals. In our case this subspace has finite
dimension and is coordinatized by a finite number of generalized
couplings.
We make the ansatz
\be\label{3.1}
\Gamma_k[\phi;g,\psi]=\Gamma_k^L[\phi;g]+
\Gamma_k^\psi[\phi;g,\psi]
                       - \tilde\kappa_k^2 I[g]/
 32\pi
                                               \ee
where
\bear  \label{3.2}
\Gamma_k^{\rm L}[\phi;g] & =&  \frac{\kappa^2_k}
{8\pi}\int d^2x\sqrt g\Bigl\{\zeta_k(\phi)D_\mu
\phi D^\mu\phi+\omega_k(\phi)R+v_k(\phi)\Bigr\}     \\
   \label{3.3}
\Gamma_k^\psi[\phi;g,\psi] &  =&  \frac{1}{16\pi}\sum_i
\left(\frac{m_{ik}}{\alpha_{ik}}\right)^2\int d^2x\sqrt g\
\psi_i\ \exp[2\alpha_{ik}\kappa_k\phi]\ear
The running parameter $\kappa_k$ is defined as the coefficient
of the $\phi R$-term
by the convention
$
 (\partial \omega_k/\partial\phi)(0)=1$.
For the main part of our analysis in ref. \cite{rwli} we assumed the
more restrictive ansatz
\be\label{3.2b}
v_k(\phi)=\frac{m_k^2}{2\alpha^2_k\kappa^2_k}\exp [2\alpha_k\kappa_k
\phi],\quad
\omega_k(\phi)=\phi,\quad \zeta_k(\phi)=\zeta_k\ee
In this case $\Gamma_k^L$ is parametrized by
four functions of $k$, namely $\kappa_k, \zeta_k$,
$m_k$ and $\alpha_k$, which
have to be determined from
the evolution equation along with
$m_{ik},\alpha_{ik}$ and
$ \tilde\kappa_k$.

Before we can embark on solving the projected RG equation we
must analyze the constraints which the Ward identities impose
on the allowed initial points $\Gamma_\Lambda$.
By inserting the above ansatz into (\ref{2.22})
and extracting the relevant field monomials from ${\cal A}_k$
in the limit $k=\Lambda\to\infty$, one finds
\be\label{3.21a}
\zeta_\Lambda=1, \quad
3\kappa_\Lambda^2=25-c+\tau , \quad
3\tilde\kappa_\Lambda^2=26-c\ee
In the classical action $S_L$ the coefficient corresponding to
$3\kappa_\Lambda^2$ is $c[S_L]=3\kappa^2=26-c$. For the translation
invariant quantization $(\tau=1)$ these values coincide, but
for the Weyl invariant case $(\tau=0)$ we have to start
from a different value, $3\kappa_\Lambda^2=25-c$. Therefore
$\Gamma_\Lambda^L$ cannot coincide with $S_L$.
We also obtain a relation which determines the
initial value $\alpha_\Lambda$ in terms of $\kappa_\Lambda$,
$
\alpha_\Lambda\kappa_\Lambda=1+2\alpha^2_\Lambda$.
With $\tau=0$ for the Weyl-invariant measure we have
the two solutions
\be\label{3.24}
\alpha_\Lambda=\frac{1}{4\sqrt3}\left[\sqrt{25-c}\pm\sqrt{1-c}\right]\ee
with the perturbative branch
corresponding to the minus sign.
They are real only for $c\leq 1$ and $c\geq25$. In the following
we restrict ourselves to $c\leq1$.
There is a similar equation for
$\alpha_{i\Lambda}$~which is most conveniently expressed in terms
of the gravitationally dressed scaling dimension relative to the area
operator, $\Delta_i(k)\equiv1-\alpha_{ik}/\alpha_k$.
For the
perturbative branch
and for $\tau=0$~it reads
\be\label{3.28}
\Delta_i(\Lambda)=\frac{\sqrt{1-c+24\Delta^0_i}
-\sqrt{1-c}}{\sqrt{25-c}-\sqrt{1-c}}
\ee
The r.h.s. of this expression is precisely what appears in the famous
KPZ-formula \cite{kpz} for the scaling dimension of $\psi_i$ in presence
of quantized gravity \cite{dk}. However, in the renormalization group
framework the properties of the quantum theory are obtained
for $k\to0$ rather than at $k=\Lambda$. Therefore (\ref{3.28})
is not the KPZ-formula yet. It obtains only provided we can
show that $\Delta_i(0)=\Delta_i(\Lambda)$.

\section{The RG Evolution } 
 The next step is to derive the
coupled system of ordinary differential equations for the functions
$\alpha_k,m_k,\cdots$. Upon inserting the truncation (\ref{3.1}) into the
renormalization group equation (\ref{2.5}) we have to evaluate
\be\label{4.1} \partial_t\Gamma_k=
         {\rm Tr}\biggl[\Bigl(\{1-\eta_k/2\} k^2C+D^2C'\Bigr)
  \Bigl(-D^2+k^2C+E_k\Bigr)^{-1}\biggr]
                       \ee
Here
$
\eta_k\equiv  -\partial_t\ln
(\kappa_k^2\zeta_k),\ C\equiv C(-D^2/k^2)$~and
$E_k=\tilde m_k^2\ \exp[2\alpha_k\kappa_k\phi]+\cdots $
with
$   \tilde m_k^2 \equiv
       m^2_k/ \zeta_k $.
By appropriate derivative and curvature expansions of the functional
trace in (\ref{4.1}) one can extract its contribution proportional
to the field monomials present in the truncation ansatz.
By comparing the coefficients of these monomials on both sides of the
RG equation one finds the set of ordinary differential equations for the
coefficient functions (generalized couplings).
From the term $ \int d^2x\sqrt gR\omega
_k(\phi)$, say, one obtains
(omitting the subscripts $k$)
\be\label{4.6a}
\partial_t[\kappa^2\omega(\phi)]=\frac{1}{3}\frac{k^2}{k^2+\tilde
m^2\exp(2\alpha\kappa\phi)}\ee
This equation shows a rather typical feature which occurs for the
other couplings as well and which is crucial for the understanding
of the RG flow in Liouville theory: depending on the relative
magnitude of the two terms in the denominator of (\ref{4.6a})
either $k^2$~or the $\phi$-dependent Liouville mass term acts as
the relevant infrared cutoff. In the latter case the couplings do not run 
any longer as a function of $k$. Let us define the $k$-dependent critical 
value $\phi_c$~by the condition that these two terms are
equal at $\phi = \phi _c$:
\be\label{4.7a}
\phi_c(k)\equiv\frac{1}{2\alpha_k\kappa_k}\ln\frac{k^2}{\tilde
m^2_k}\ee
We see that $\kappa^2\omega(\phi)$
changes at a constant rate $\partial_t[\kappa^2\omega]
=\frac{1}{3}$ for $\phi\ll\phi_c$~($k$~large) whereas the evolution
stops for $\phi\gg\phi_c$
($k$~small) due to an exponential suppression factor:
\be\label{4.7c}
\partial_t\left[\kappa^2\omega(\phi\gg\phi_c)\right]=\frac{1}{3}
\exp[-2\alpha\kappa(\phi-\phi_c)]\ee
The subtle aspect of this decoupling phenomenon is that in different
regions of field space the threshold separating the two regimes
is crossed at different values of $k$.

For a detailed discussion of the evolution equations for the
various couplings we have to refer to \cite{rwli} where also
the approximations are described, which went into their solution.
Here we only quote the result for $\Gamma_k$~in the physical
limit $k\to 0$. One finds the $k$-independent functional
\bear
\label{lim}
\Gamma_0 &=&
\frac{26-c}{24\pi}\int d^2x\sqrt{g}       \bigg\{
 D_\mu\phi D^\mu\phi+ R\phi+\frac{\bar m^2}
{2}e^{2\phi}   \nonumber  \\
& &\ \ \ \ \    +  \sum_i \frac{\bar m^2_i}{2}  \psi_i  \exp[2(1-
\Delta^0_i)\phi]
 \bigg\}    -\frac{26-c}{96\pi} \  I[g]  \ear
Remarkably enough, up to modified mass parameters $\bar m$~and
$\bar m_i$~this is precisely the classical action.
In particular, the value of the coefficient $3\kappa_k^2$~has increased
from $25-c$~at the initial point $k=\Lambda$~to $26-c$~at the final
point $k=0$ \footnote{We set $\tau = 0$~from now on.}.
Here we encounter the slightly unusual situation
that even during an infinitely long running ``time'' ($\Lambda\to
\infty$) a certain coupling changes by a finite amount only
\footnote{A similar finite renormalization was found for
          the Chern-Simons parameter in $d=3$ \cite{cs} and for the
          $\theta$-parameter in $d=4$ \cite{tch}.}.
The same remark also applies to $\alpha_k$~which changes in such a way
that $\alpha_0\kappa_0 = 1$, i.e., for $k\to 0$~we recover
the classical Liouville potential of (\ref{1.7}).
The fact that the effective potential equals the classical one
can be proven \cite{rwli} on the basis of the rather general
truncation (\ref{3.2}); it is not necessary to make the
ansatz (\ref{3.2b}).
Likewise the exponentials multiplying the $\psi_i$'s return to their
classical form as $k$~approaches zero. One obtains
$ \lim_{k\to0}\alpha_{ik}\kappa_k=
                      1-\Delta_i^0        $.
This implies that the gravitationally dressed scaling dimensions
equal the classical ones,
$
\Delta_i(k=0)=
 \Delta_i^0$.
Even though it seems that our truncation is too simple to
reproduce the KPZ formulas in the infrared, one can verify that
(within the approximations which were made in solving the evolution
equations) the resulting RG trajectory is perfectly consistent
with the Ward identities \cite{rwli}.
In particular, $\Gamma_0$~is easily seen to describe a conformal
field theory with total central charge zero.

\section{The UV-Fixed Point}
Let us look at the space of 2-dimensional field
theories with one scalar field as a whole now. Typically
a critical scalar theory is governed by a fixed
point which is IR-unstable in one or several
directions. The flow starts then for $k=\Lambda$ in the
immediate vicinity of this fixed point and remains there
for a long ``running time''. Depending on the value of
$\phi$ the running stops at a certain point,
and $\Gamma_k[\phi]$ becomes
independent of $k$ for $k\to0$. Generically there is also
a region in $\phi$ where the running never stops. In the
critical ${\bf Z}_2$-symmetric $\phi^4$-theory
(Ising model) this region shrinks to one point --
the potential minimum -- as $k\to0$. In our case this region
is $\phi<\phi_c(k)$ and it moves to $\phi\to-\infty$ as
$k\to0$.

We search for fixed points of the renormalization group
flow in the subspace
of action
functionals of the form
(``local potential approximation'')
\be\label{7.1}
\Gamma_k[\phi;g]=Z_k\int d^2x\sqrt g\left\{\frac{1}{2}
D_\mu\phi D^\mu\phi+Rf_k(\phi)+k^2u_k(\phi)\right\}\ee
Here $f_k$ and $u_k$ are arbitrary dimensionless functions.
Hence the condition for a fixed point is
simply
$\partial_t u_k=\partial_t f_k=\partial_tZ_k=0$.
By inserting (\ref{7.1}) into (\ref{2.5}) we obtain the following
equation for the fixed point potential $u_*(\phi)$:
\be\label{7.3}
8\pi Z_*\ u_*(\phi)=\int^\infty_0dy\frac
{C(y)-yC'(y)}{y+C(y)+u_*''(\phi)}\ee
It is not difficult to see \cite{rwli} that a generic solution
to this equation is a periodic function of $\phi$.
Solutions of this type are not related to Liouville theory
and represent different universality classes.
The only solution which is not periodic is the constant one,
$ u_{*} (\phi) \equiv u_{*} $.
We are now going to show that Liouville theory can be
understood as the perturbation of this ``Gaussian'' fixed point by a
relevant operator. When $k$~is lowered from infinity down to zero, this
operator drives the theory from the IR unstable Gaussian fixed
point to a different IR stable one
which represents {\it quantum} Liouville theory.
In order to linearize the RG flow in the vicinity of the Gaussian
fixed point
we write
$
u_k(\phi)=u_*+\varepsilon\delta u_k(\phi)   $
with $(t\equiv\ln [k/\Lambda])$
\be\label{7.9}
\delta u_k(\phi)=e^{-\gamma t}\ \Upsilon(\phi)+\sum_i
\ e^{-\gamma_it}\ e^{2\Delta^0_it}
\ \psi_i\  \Upsilon_i(\phi)\ee
Here $\Upsilon(\phi)$ and $\Upsilon_i(\phi)$ are the eigenvectors of the
linearized evolution operator with the eigenvalues
$\gamma$ and $\gamma_i$, respectively.
By expanding the evolution equation to first order in $\varepsilon$
and solving the resulting linear equation
one obtains for the potential near the fixed point:
\be\label{7.14}
               k^2\delta u_k(\phi)  \propto  m_\Lambda^2
  \left[\left(
 \Lambda /k \right)^{4\alpha^2_*}e^{2\alpha_*\kappa_*\phi}+
\sum_i\psi_i\left(
 \Lambda /k \right)^{4\alpha^2_{i*}}e^{2\alpha_{i*}\kappa_*\phi}
\right]\ee
This result is universal in the sense that it does not depend
on the form of the cutoff function $C(\cdot)$.
Now we understand what is so special about the exponential
interaction potentials: they are precisely the eigenvectors of the
linearized renormalization group flow at the Gaussian fixed point.
For $\alpha_*,\alpha_{i*}$ real,
the perturbation grows for
decreasing $k$ and this fixed point is IR-unstable therefore. Hence we
can identify $\alpha_*,\alpha_{i*}$ and $\kappa_*$ with the
UV parameters $\alpha_\Lambda,\alpha_{i\Lambda}$ and 
$\kappa_\Lambda$ whose values are fixed by the Ward identities.
So far we discussed only the potential term. If we also take the
fixed point condition for $f_k$~into account we find that the
fixed point action is of the Feigin-Fuks type:
\be
\label{ff}
\Gamma_*  =
\frac{25-c}{24\pi}\int d^2x\sqrt{g}       \big(
 D_\mu\phi D^\mu\phi+ R\phi + {\rm const}
 \big)         \ee
This theory is conformally invariant with
$c[\Gamma_*]=25-c$.

Note that the area
operator is required to be an $(1,1)$ operator in the {\it quantum}
theory, i.e., for $k\to0$, but not necessarily at $k=\Lambda$. In the
UV it is not marginal and this is what drives the system away from
the UV-fixed point. In fact, the area operator responds
to a conformal reparametrization
$z\to z'(z)$, $d{s'}^2=|dz'/dz|^2ds^2,\phi'=\phi-\frac{1}{2}\ln
|dz'/dz|^2$~according to
\be\label{7.15}
(\sqrt ge^{2\alpha_\Lambda\kappa_\Lambda\phi})'
=\left|dz'/dz\right|^{-4\alpha^2_\Lambda}\sqrt ge^{2\alpha
_\Lambda\kappa_\Lambda\phi}\ee
This is a direct consequence of the Ward identity
$
\alpha_\Lambda\kappa_\Lambda=1+2\alpha^2_\Lambda$.

\section{Conclusion}
Let us summarize the results obtained from
the Ward identities, the evolution equation
and the fixed point equation. The overall picture which emerges is that
of a crossover phenomenon from one fixed point to another.
The renormalization group trajectory
starts for $k=\Lambda$ at a conformal theory with central charge
$3\kappa^2_\Lambda=25-c$. This
initial point, approached for $\Lambda\to\infty$, is
a Feigin-Fuks free field action
$\Gamma_*$. The
perturbations at this fixed point are governed by eigenvectors
of the linearized RG flow, $\Upsilon(\phi)$ and
$\Upsilon_i(\phi)$, which have an
exponential dependence on $\phi$. Adding such a perturbation
to $\Gamma_*$ we obtain the Liouville action. The
Weyl--Ward identities fix the free parameters
in this action (except for $m_\Lambda$). For these initial values
the perturbation is a relevant operator which
drives the system away from the IR-unstable fixed
point. For $k\to0$, $\Gamma_k^L$ approaches an IR-stable
fixed point.
The Ward identity
guarantees that $\Gamma_{k\to0}^L$ is a conformal field theory
with $c[\Gamma_{k\to0}]=26-c$. Within our truncation we find
that $\Gamma_0^L$ equals the classical Liouville action.
We expect that this is a very good approximation (and
perhaps even exact) for the low momentum part of the effective
action. An improved truncation should, for instance,
reproduce the KPZ dimensions
at the {\it final} point $k=0$.
It remains slightly mysterious why the present truncation yields
precisely the KPZ scaling dimensions at the
{\it initial} point of its RG trajectory.

\end{document}